\RequirePackage{fix-cm}
\documentclass[smallextended]{svjour3}       
\smartqed  
\usepackage{graphicx}
\usepackage{url}
\usepackage{color}
\journalname{Experimental Astronomy}

\begin{document}

\title{Background simulations for the Large Area Detector onboard LOFT}

\author{Riccardo~Campana      
	\and Marco~Feroci
	\and Ettore~Del~Monte
	\and Teresa~Mineo
	\and Niels~Lund
	\and George~W.~Fraser
	\thanks{On~behalf~of~the~LOFT~collaboration.}
}

\authorrunning{R. Campana et al.} 

\institute{R. Campana, M. Feroci, E. Del Monte \at
              INAF/IAPS, Via Fosso del Cavaliere 100, I-00133, Roma, Italy \\
              and INFN/Sezione di Roma 2, viale della Ricerca Scientifica 1, I-00133, Roma, Italy. \\
              \email{riccardo.campana@inaf.it}             \\
             \emph{Present address of R. Campana:} \\ INAF/IASF-Bologna, via Gobetti 101, I-40129, Bologna, Italy.\\
           T. Mineo \at
              INAF/IASF-Palermo, Via Ugo La Malfa 153, I-90146, Palermo, Italy. \\
             N. Lund \at
             DTU-Space, Elektrovej Bld. 327, DK-2800, Kgs. Lyngby, Denmark. \\
             G. Fraser \at
             Space Research Centre, Dept. of Physics and Astronomy, University of Leicester, LE17RH, Leicester, UK.
}

\date{Received: date / Accepted: date}

\maketitle

\begin{abstract}
The Large Observatory For X-ray Timing (LOFT), currently in an assessment phase in the framework the ESA M3 Cosmic Vision programme, is an innovative
medium-class mission specifically designed to answer fundamental questions about the behaviour of matter, in the very strong gravitational and magnetic fields
around compact objects and in supranuclear density conditions. Having an effective area of $\sim$10 m$^2$ at 8 keV, LOFT will be able to measure with high
sensitivity very fast variability in the X-ray fluxes and spectra. 
A good knowledge of the in-orbit background environment is essential to assess the scientific
performance of the mission and optimize the design of its main instrument, the Large Area Detector (LAD).  
In this paper the results of an extensive Geant-4 simulation of the instrument will be discussed, showing  the main contributions to the background and the design 
solutions for its reduction and control.
Our results show that the current LOFT/LAD design is expected to meet
its scientific requirement of a background rate equivalent to 10 mCrab in 2--30 keV,
achieving about 5 mCrab in the most important 2--10 keV energy band. 
Moreover, simulations show an anticipated modulation of the background rate as small as 
10\% over the orbital timescale. 
The intrinsic photonic origin of the largest background component also allows for an efficient modelling,
supported by an in-flight active monitoring, allowing to predict systematic residuals significantly better than the requirement of 1\%, and actually 
meeting the 0.25\% science goal.
\keywords{X-ray astronomy \and Instrumental background \and Montecarlo simulations}
\PACS{07.85.Fv \and 07.87.+v \and 95.40.+s}
\end{abstract}

\section{Introduction}\label{s:intro}
Astrophysical observations in the X-ray domain are often dominated by the background.
In order to assess and maximize the scientific performance of a satellite-borne instrument during its design phase, is essential to have an early estimation of the background level and to identify its origin.
This will allow to drive the optimization of the experiment design itself.
Moreover, a reliable estimation of the background level is important to assess the scientific objectives and performance of the mission.
The most convenient and consolidated way to compute such a background is by means of a Monte Carlo simulation, using e.g.  the Geant-4 toolkit \cite{agostinelli03}, in which the interactions of the space environment with a somewhat simplified mass model of the spacecraft and of the instruments are sampled and studied.
A correct evaluation of the radiation environment surrounding the experiment is of course the most important ingredient in the background simulation.

In this paper we will show Monte Carlo simulations of the instrumental background for the Large Area Detector (LAD) instrument onboard the proposed LOFT (Large Observatory for X-ray Timing) satellite \cite{feroci11,zane12}, that 
is a candidate for the third slot of medium-class missions (M3) of the Cosmic Vision 2015--2025 programme of the European Space Agency, with a possible launch in the 2022--24 timeframe.

\begin{figure}[htbp]
\centering
\includegraphics[scale=0.55]{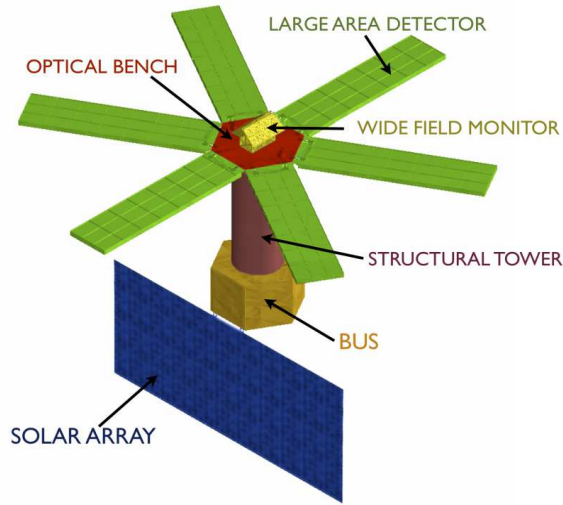}
\caption{A sketch of the LOFT satellite and its instruments.}
\label{loftoverview}
\end{figure}

LOFT (Figure \ref{loftoverview}) has been specially designed to answer the ``Fundamental Question" \#3.3 of the Cosmic Vision programme, \emph{Matter under extreme conditions}.
In the aim of investigating the behaviour of the matter in the most extreme physical conditions that can be found in the gravitational and magnetic fields around neutron stars and black holes, LOFT will study the rapid variability of the X-ray emission from these objects, both in the spectral and temporal domains.
Moreover, LOFT will be an observatory for virtually all classes of bright X-ray sources: among them, X-ray bursters, high mass X-ray binaries, cataclysmic variables, magnetars and active galactic nuclei.

The main experiment onboard LOFT is the LAD \cite{zane12} (Sect. \ref{s:lad}), a collimated instrument with a collecting area 20 times bigger than its immediate predecessor (PCA, the Proportional Counter Array onboard the Rossi X-ray Timing Explorer, RXTE), and based on large-area Silicon Drift Detectors (SDDs) coupled to lead-glass Micro Channel Plate (MCP) collimators.
In order to survey a large fraction of the sky simultaneously, and to trigger follow-up observations with the main instrument, LOFT will host also a coded-mask Wide Field Monitor (WFM, \cite{brandt12}). 
This latter instrument, sensitive in the same energy range of LAD, will use basically the same SDD detectors, with some design differences to optimize the detector response for imaging \cite{campana11,evangelista12,donnarumma12}.

The background level required to satisfy the LOFT scientific objectives, in particular regarding the relatively faint (1--10 mCrab) sources like most Active Galactic Nuclei (AGNs) and some accretion-powered X-ray pulsars, is $\le$10 mCrab for the 2--30 keV LAD standard band.
Being a collimated instrument, the LAD will not measure simultaneously and independently the source flux and the background. 
While some collimated instruments (e.g. BeppoSAX/PDS and RXTE/HEXTE, both at hard X-rays) employ a ``rocking'' strategy, where the same instrument unit alternates observations of the source and of the background, others (e.g. RXTE/PCA, at soft X-rays, similar to LAD) rely on a background model obtained by means of dedicated blank-field pointings and by using onboard ratemeters \cite{jahoda06}.
As we will discuss in the following, LAD will adopt an approach similar to PCA. 
To ensure the fulfillment of the scientific objectives, in particular for low flux sources, an accurate estimation and modelling of the background level for this instrument is of particular importance. The background variations that have a significant impact on the scientific performance should be minimized, and controlled to a high degree of accuracy. 
The LOFT science case resulted in a requirement on background knowledge at 1\%, with a goal at 0.25\%.
As a comparison, RXTE/PCA reached a background residual systematic level of $\sim$0.5\%--1\% \cite{jahoda06,shaposhnikov12}, while BeppoSAX/PDS of $\sim$1\% \cite{frontera97}.

This paper is structured as follows. In Section \ref{s:lad} we describe the LAD instrument. In Section \ref{s:bkgsources} we introduce the space environment for LOFT and discuss the main contributions to the instrumental background, while in Section \ref{s:massmod} the main features of the LAD Monte Carlo mass model and simulator are described. The simulation results are shown in Section \ref{s:results}, and in Section \ref{s:conclusions} we draw our conclusions.

\section{The Large Area Detector}\label{s:lad}

\begin{figure}[htbp]
\centering
\includegraphics[width=\textwidth]{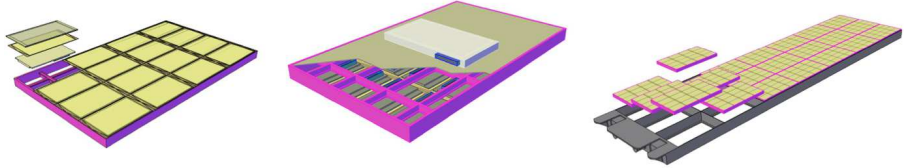}
\caption{\emph{Left}: a LAD module, showing the mounting of the collimator, SDD and the FEE. \emph{Center}: Back-side view of the module, showing the radiative surface and the back-end electronics. \emph{Right}: a LAD Detector Panel.}
\label{f:panel}
\end{figure}

The Large Area Detector is an array of 2016 individual large-area Silicon Drift Detectors, sensitive to the 2--30 keV emission (with a extended 30--80 keV band)  collimated in a $\sim$1$^\circ$ field of view by means of a lead-glass microcapillary collimator plate. 
The instrument design is modular: an array of 4$\times$4 SDDs and their front-end electronics (FEE) are placed in a \emph{module}. 
Each module has its independent back-end electronics (MBEE) that powers and configure the FEE, manages the event readout and monitors the detector health.
A \emph{panel} is composed of an array of 21 modules, placed on a support grid (Figure \ref{f:panel}).

The current LAD design by the LOFT consortium envisages 6 panels (each about 350 cm $\times$ 90 cm), that stems from the satellite bus and are deployed once the spacecraft reaches its orbit (Figure \ref{loftoverview}).
The collimators, made of lead glass, have the same footprint as the SDDs (about 11 cm $\times$ 7 cm), a thickness of 6 mm, and 100 $\mu$m wide square holes separated with 20 $\mu$m thick walls (and thus having an \emph{open area ratio}, $f_\mathrm{OAR}$, of $\sim$70\%).
An aluminium frame couples the collimators with the detectors frame and ensures the tight alignment constraints.

The working principle of the SDDs is the following: a photon is absorbed in the depleted silicon bulk, and produces a cloud of electrons whose number is proportional to the incident energy. 
The cloud drifts towards an array of read-out anodes, by means of a high-voltage field applied between the middle section and the two detector edges, hosting two independent rows of anodes.
During the drift, the cloud size increases due to diffusion. The number of anodes that will read-out the charge cloud thus depends on the absorption point distance and anode pitch \cite{campana11}. 
For the LAD detectors, that have a pitch of 970 $\mu$m, the charge cloud originating from the absorption of a X-ray photon in the 2--80 keV band is at most 0.5 mm (FWHM) thus read-out by 1 or 2 anodes, depending on the impact point and the energy. At 6 keV, for example,  $\sim$40\% of the events are collected by one anode and $\sim$60\% by two anodes.
The event multiplicity is read-out and transmitted to the ground, enabling a selection of ``single'' and ``double'' events, with different energy resolutions.

\section{Sources of background}\label{s:bkgsources}
In this section we  discuss the main sources of the LOFT background. We  describe the environment encountered in the baseline low-Earth orbit, with its hadronic, leptonic and photon components. Moreover, we describe the internal source of background, i.e. the natural radioactivity of the collimator material.

\subsection{The LOFT orbit}
The LOFT orbit requirement is at an altitude of $h = 600$ km with an inclination of 5$^{\circ}$. 
Likewise, lower-altitude orbits (down to 550 km) and/or inclinations (down to 0$^{\circ}$) are considered, with small improvements for what regards the conclusions about the background rate and properties. The lower altitude and inclinations, however, impacts on the long-term radiation damage on the detectors.
For these orbits, the geomagnetic latitude range is usually  $\theta_{M} < 0.3$ radians.

\begin{figure}[htbp]
\centering
\includegraphics[width=0.9\textwidth]{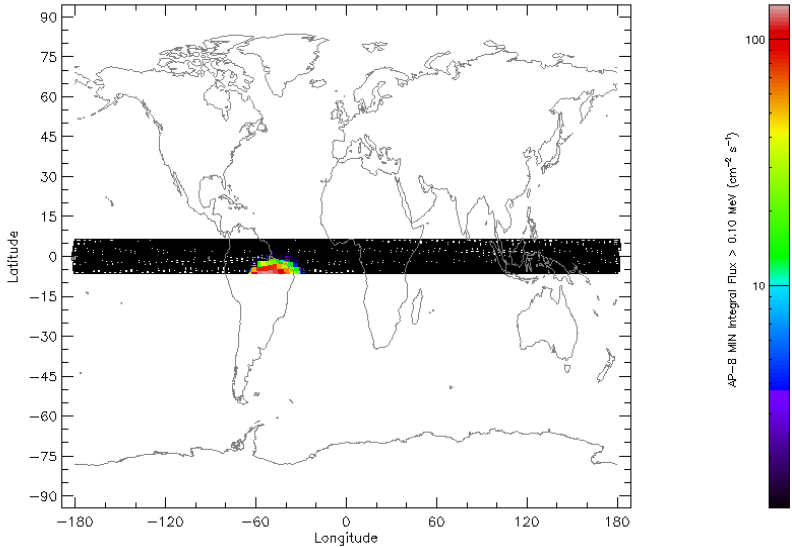}
\caption{Values of the trapped proton intensity (AP8MIN model) for the LOFT baseline orbit. Data obtained from the SPENVIS software (\texttt{http://www.spenvis.oma.be}). The South Atlantic Anomaly is well apparent above South America.}
\label{f:saa}
\end{figure}

In the current work we neglected the possible background component due to the activation of the instrument and satellite materials. Based on the arguments below, we estimated it to be a second-order effect. 
The South Atlantic Anomaly, a region of trapped high-energy protons and electrons, is grazed by the LOFT baseline orbit only in its outermost regions (Figure \ref{f:saa}), with a shallower passage for lower altitudes and inclinations. 
Therefore, the effect on background due to the activation of materials by this intense radiation is expected to be negligible with respect to other sources, as confirmed by preliminary evaluations and heritage from past missions in similar orbits (e.g., BeppoSAX and AGILE).
It should also be noted that in contrast to previous experiments, the LAD instrument has a very light structure per unit volume and the relatively small spacecraft assembly is seen at a small viewing angle, due to the tower supporting the LAD panels (see Figure~\ref{loftoverview}). In addition to that, as for example observed by Swift/XRT \cite{pagani07}, Silicon detectors show marginal activation effects in LEO, even when operated as focal plane instruments (more local mass) and on more inclined orbits (up to $\sim$20$^\circ$).
However, given our goal of controlling the LAD background to the highest possible accuracy, we plan to study this aspect in great detail in a future work.

\subsection{Primary cosmic rays}\label{s:primarycr}
Following the approach of Mizuno et al. \cite{mizuno04}, we assume that the primary cosmic ray spectrum is expressed as $F_U(R)$, a function of the particle magnetic rigidity, $R = pc/Ze$, where $Z$ is the atomic number and $p$ the particle momentum.

The full spectrum at a given location in the magnetosphere and a given phase of the solar cycle will be given by:
\begin{equation}
F(E) = F_U(E + Ze\phi) \times F_M(E, M, Z, \phi) \times C(R, h, \theta_{M})
\end{equation}
where $M$ and $Ze$ are the mass and charge of the particle, $E$ its kinetic energy, $\phi$ is a solar modulation factor, $h$ is the orbit height and $\theta_{M}$ is the geomagnetic latitude.

The second term, that includes the effects of the solar modulation on the cosmic ray particles, is given by \cite{gleeson68}:
\begin{equation}
F_M(E, M, Z, \phi) = \frac{(E+Mc^{2})^{2} - (Mc^{2})^{2}}{(E+Ze\phi+Mc^{2})^{2} - (Mc^{2})^{2}}
\end{equation}
where the \emph{solar modulation potential} $\phi$ is given by:
\begin{equation}
 \phi = 0.55  \mbox{\, GV \, \, at solar minimum}
 \end{equation}
\begin{equation}
 \phi = 1.10 \mbox{\, GV \, \,  at solar maximum}
\end{equation}

The geomagnetic cutoff function is given, for vertically incident particles, by \cite{mizuno04}:
\begin{equation}\label{eq:geomag_cutoff}
C (R, h, \theta_{M}) = \frac{1}{1 + (R/R_{\mathrm{cut}})^{-r}}
\end{equation}
in which the \emph{cutoff rigidity} is obtained from the St\"ormer equation \cite{smart05}:
\begin{equation}
R_{\mathrm{cut}} = 14.5 \times \left( 1 + \frac{h}{R_{E}} \right)^{-2} \cos^{4} \theta_{M} \mbox{\, GV}
\end{equation}
where $R_{E} = 6371$ km is the Earth radius. Moreover, the $r$ exponent in Eq. \ref{eq:geomag_cutoff} is:
\begin{equation}
 r = 12  \mbox{\, for protons}
 \end{equation}
\begin{equation}
 r = 6  \mbox{\, for electrons and positrons}
\end{equation}
For the 600 km, 5$^\circ$ inclination LOFT orbit, the cutoff rigidity therefore is found to be in the range $R_{\mathrm{cut}} \sim $ 10.1--12.1 GV.

\begin{figure}[htbp]
\centering
\includegraphics[width=0.9\textwidth]{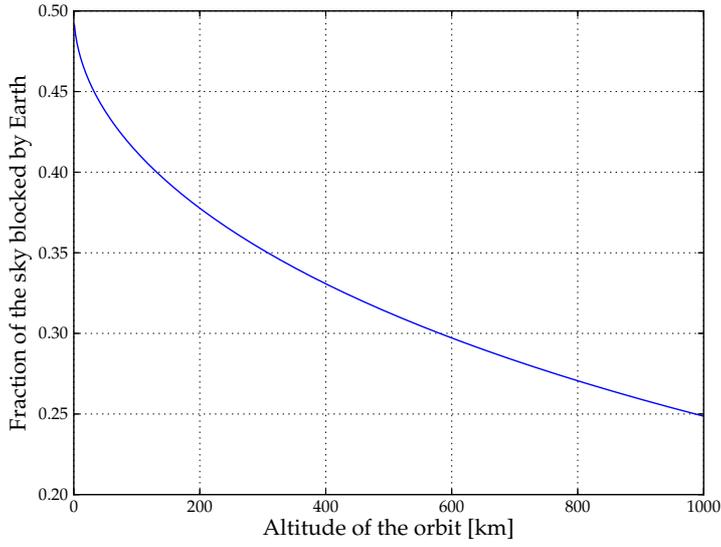}
\caption{The fraction of the sky blocked by the Earth, for a given orbit altitude $h$.}
\label{f:sky}
\end{figure}

Since we are interested mostly in the average background fluxes, we discard at the present stage the east-west effect for which particles coming from different directions have different cutoff rigidities.
We therefore assume that the general primary spectrum, now expressed as:
\begin{eqnarray}
F(R) &= F_U(E + Ze\phi) \times  \frac{(E+Mc^{2})^{2} - (Mc^{2})^{2}}{(E+Ze\phi+Mc^{2})^{2} - (Mc^{2})^{2}} & \nonumber\\
& \times \frac{1}{1 + (R/R_{\mathrm{cut}})^{-r}}
\end{eqnarray}
has an \emph{uniform} angular distribution with respect to the zenith angle, up to the Earth horizon, i.e. for zenith angles $\theta$ from 0$^{\circ}$ to $\theta_{\mathrm{cut}}$, where the latter is given by:
\begin{equation}
\theta_{\mathrm{cut}} = \arcsin \left( \frac{1}{\frac{h}{R_{E}} + 1} \right)
\end{equation}
For $h = 600$ km, $\theta_{\mathrm{cut}} \sim 114^{\circ}$. 
The solid angle subtended by the Earth at an altitude $h$ is given by:
\begin{equation}
\Omega_{E} =  2 \pi \left(1 - \frac{1}{R_{E}/h+1}\sqrt{1+\frac{2R_{E}}{h}} \right) 
\end{equation}
and therefore the accessibile sky subtends a solid angle $\Omega = 4\pi - \Omega_{E}$ (Figure~\ref{f:sky}).
For $h  = 600$ km, the Earth blocks about 30\% of the sky.

\subsubsection{Primary protons}
The unmodulated value of the cosmic ray electron spectrum is given by the BESS \cite{sanuki00} and AMS measurements \cite{alcaraz00a}:
\begin{equation}
F_U(E) = 23.9 \times \left[ \frac{R(E)}{\mathrm{GV}} \right]^{-2.83} \mbox{\, particles m$^{-2}$ s$^{-1}$  sr$^{-1}$  MeV$^{-1}$}
\end{equation}

\begin{figure}[htbp]
\centering
\includegraphics[width=0.9\textwidth]{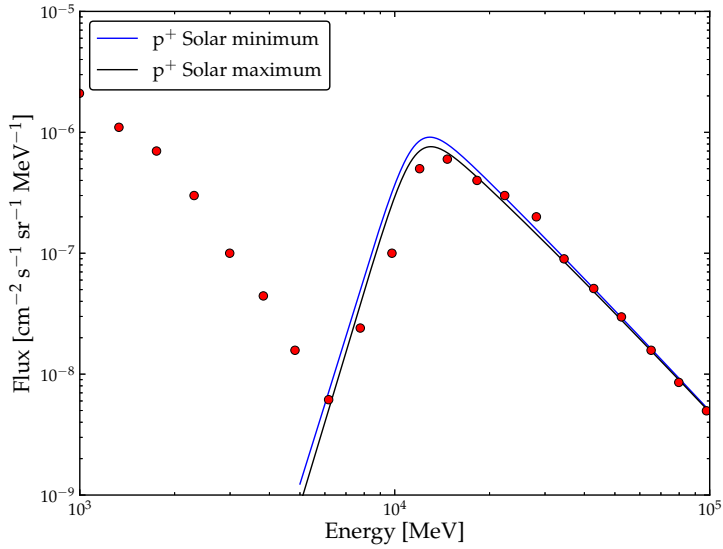}
\caption{Model spectrum of the primary protons  for the LOFT orbit. The red data points are the AMS-01 measurements of the proton spectrum, primary and secondary, for the Shuttle orbit (380 km) and low magnetic latitude ($\theta_{M} < 0.2$), from \cite{alcaraz00a}.}
\label{f:protons}
\end{figure}

In Figure \ref{f:protons}  the primary proton spectra for different values of the solar modulation and for the LOFT orbit is reported.
The variation in flux between solar minimum and maximum, at the peak flux, is $\sim$15\%.

\subsubsection{Primary electrons and positrons}

The unmodulated value of the cosmic ray electron spectrum is given by \cite{webber83,moskalenko98,mizuno04}:
\begin{equation}
F_U(E) = 0.65 \times \left[ \frac{R(E)}{\mathrm{GV}} \right]^{-3.3} \mbox{\, particles m$^{-2}$ s$^{-1}$  sr$^{-1}$  MeV$^{-1}$}
\end{equation}

The fraction of positrons to electrons, usually given by the ratio $e^{+}/(e^{+}+e^{-})$,  is found to be rather independent of the energy \cite{golden94}, i.e. the spectrum of primary positrons has the same slope of the electron one, but a different normalization:
\begin{equation}
F_U(E) = 0.051 \times \left[ \frac{R(E)}{\mathrm{GV}} \right]^{-3.3} \mbox{\, particles m$^{-2}$ s$^{-1}$  sr$^{-1}$  MeV$^{-1}$}
\end{equation}

\begin{figure}[htbp]
\centering
\includegraphics[width=0.9\textwidth]{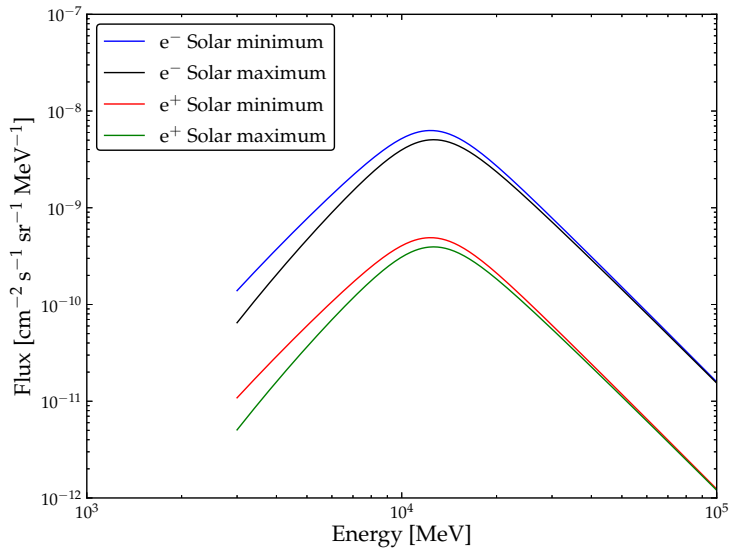}
\caption{Model spectrum of the primary electrons and positrons for the LOFT orbit.}
\label{f:electrons}
\end{figure}

In Figure \ref{f:electrons} the primary $e^{-}$ and $e^{+}$ spectra for different values of the solar modulation and for the LOFT orbit are reported. 
The difference in flux between solar minumum and maximum is about 20\% at the peak.

\subsubsection{Primary alpha particles}
\begin{figure}[htbp]
\centering
\includegraphics[width=0.9\textwidth]{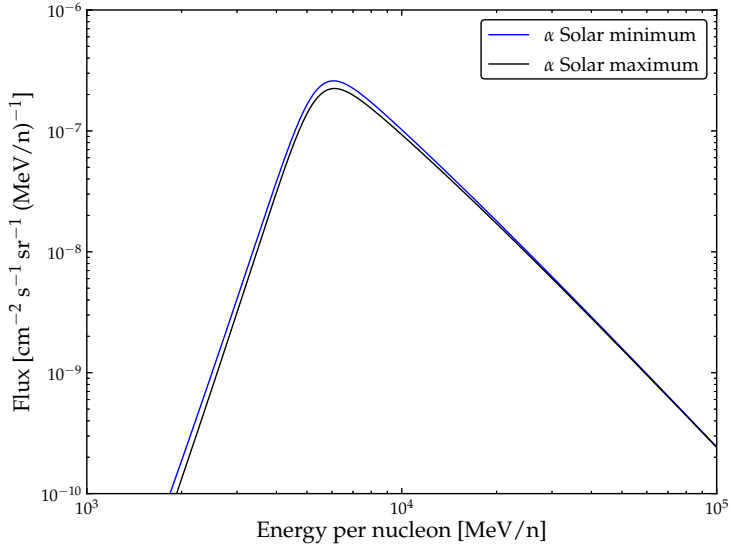}
\caption{Model spectrum of the primary helium nuclei  for the LOFT orbit.}
\label{f:alpha}
\end{figure}

We consider only the primary helium nuclei, because the contribution of heavier nuclei is negligible with respect to the other primary particles flux \cite{zuccon04}.
The unmodulated flux, as given by Mizuno et al. \cite{mizuno04} on the basis of AMS and BESS data is:
\begin{equation}
F_U(E) = 1.5 \times \left[ \frac{R(E)}{\mathrm{GV}} \right]^{-2.77} \mbox{\, particles m$^{-2}$ s$^{-1}$  sr$^{-1}$  MeV$^{-1}$}
\end{equation}
and is shown in Figure \ref{f:alpha}.

\subsection{Secondary particles}

\subsubsection{Protons}
\begin{figure}[htbp]
\centering
\includegraphics[width=0.9\textwidth]{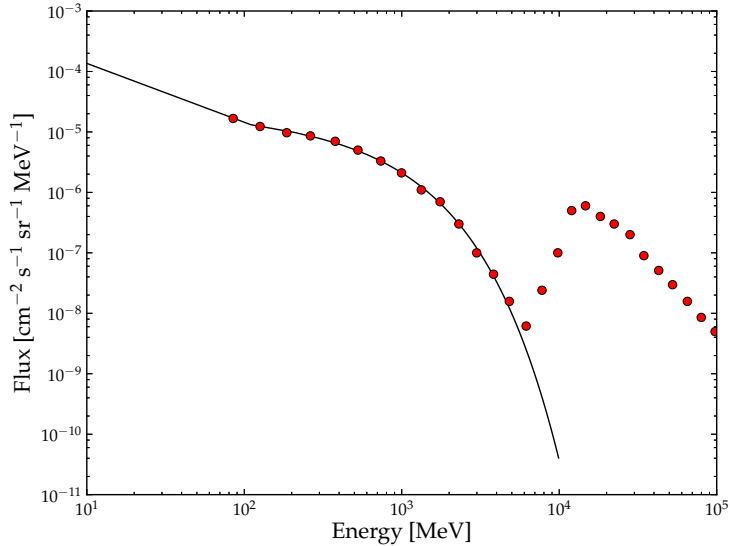}
\caption{Secondary proton spectrum. The red data points are the AMS-01 measurements of the proton spectrum, both primary and secondary, for the Shuttle orbit (380 km) and low magnetic latitude ($\theta_{M} < 0.2$, from \cite{alcaraz00a}), while the black line is the Mizuno et al. \cite{mizuno04} analytic form.}
\label{f:secondaryProtons}
\end{figure}

For the low altitude equatorial Earth orbits considered, the impinging proton spectrum outside the trapped particle belt, i.e. the South Atlantic Anomaly, consists of the primary component discussed in Sect.~\ref{s:primarycr} and of a secondary, quasi-trapped component, originating from and impacting to the Earth atmosphere (sometimes in the literature they are referred to as the ``splash'' and ``reentrant'' albedo components). AMS measurements \cite{alcaraz00a} showed that this secondary component is composed of a short-lived and a long-lived particle population, both originating from the regions near the geomagnetic equator.
Mizuno et al. \cite{mizuno04} model the secondary equatorial proton spectrum as a cutoff power-law:
\begin{equation}
F(E) = 0.123 \times  \left( \frac{E}{1 \mathrm{\, GeV}} \right)^{-0.155} e^{- \left(E/0.51\right)^{0.845}} \mbox{\, particles m$^{-2}$ s$^{-1}$  sr$^{-1}$  MeV$^{-1}$}
\end{equation}
and a power law for energies below 100~MeV:
\begin{equation}
F(E) = 0.136 \times \left( \frac{E} {100 \mathrm{\, MeV}} \right)^{-1} \mbox{\, particles m$^{-2}$ s$^{-1}$  sr$^{-1}$  MeV$^{-1}$}
\end{equation}
This spectrum is shown in Figure \ref{f:secondaryProtons} together with the AMS measurements. 
According to the measurements performed with the NINA and NINA-II instruments \cite{bidoli02}, the extrapolation below 100~MeV is likely an overestimate.

We do not consider here the soft (10 keV--1 MeV), highly directional equatorial proton population \cite{petrov08,petrov09}, 
since these particles are most efficiently stopped by all the spacecraft structures surrounding the detectors. 
Even the small fraction that impinges on the detector through the capillary holes, 
albeit significant when considering the long-term radiation damage on the SDDs, leaves a negligible background signal.

\subsubsection{Electrons and positrons}
The equatorial secondary electron spectrum can be approximated with a broken power law \cite{mizuno04}:
\begin{equation}
F(E) = 0.3 \times  \left( \frac{E}{100 \mathrm{\, MeV}} \right)^{-2.2} \mbox{\, for 100 MeV $< E < 3 $ GeV}
\end{equation}
\begin{equation}
F(E) =  0.3 \times  \left( \frac{3 \mathrm{\, GeV}}{100 \mathrm{\, MeV}} \right)^{-2.2} \left( \frac{E}{3 \mathrm{\, GeV}} \right)^{-4} \mbox{\, for $E \ge 3$ GeV}
\end{equation}
and a power law for energies below 100 MeV:
\begin{equation}
F(E) = 0.3 \times \left( \frac{E} {100 \mathrm{\, MeV}} \right)^{-1} \mbox{\, particles m$^{-2}$ s$^{-1}$  sr$^{-1}$  MeV$^{-1}$}
\end{equation}

\begin{figure}[htbp]
\centering
\includegraphics[width=.9\textwidth]{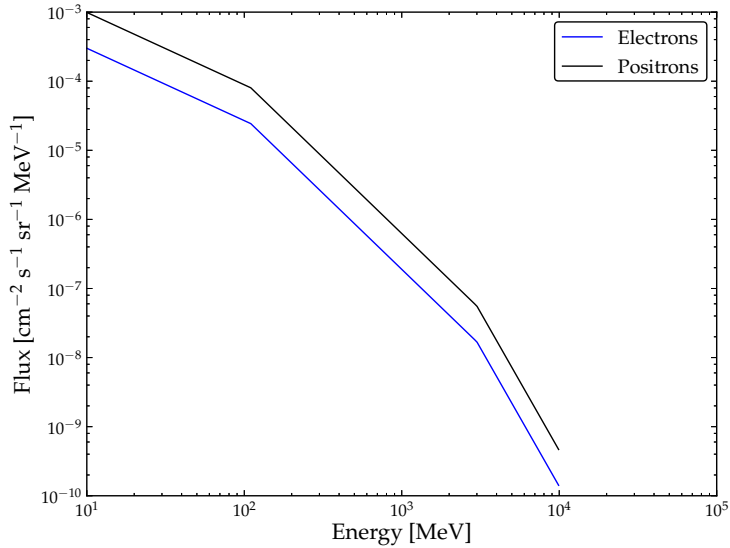}
\caption{Secondary electron and positron spectrum. Analytic form of Mizuno et al. \cite{mizuno04}.}
\label{f:secondaryElectrons}
\end{figure}

At variance with respect to the primary particles, in the geomagnetic equatorial region the positrons are predominant with respect to the electrons. The spectrum is the same, but the ratio $e^{+}/e^{-}$ is about 3.3 (Figure \ref{f:secondaryElectrons}).

\subsection{Photon background}
As we will see in the following Sections, the main contribution to the overall background for the LAD instrument is due to the high energy X-ray emission transmitted and scattered in the lead-glass collimators. 
The main contribution comes from the CXB and the Earth albedo emission.

\subsubsection{Cosmic photon background}
\begin{figure}[htbp]
\centering
\includegraphics[width=.9\textwidth]{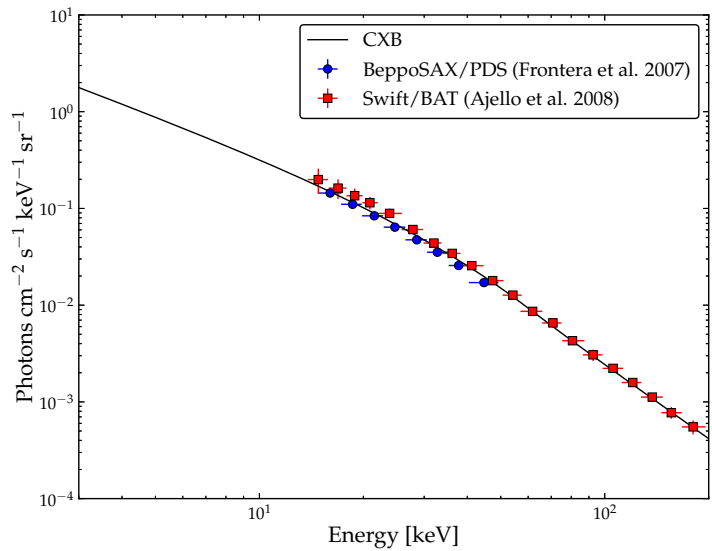}
\caption{The cosmic X-ray diffuse background as derived from HEAO-1 A4 measurements \cite{gruber99} (solid line). CXB observations performed by PDS \cite{frontera07} and BAT \cite{ajello08} are also shown.}
\label{f:cxb}
\end{figure}

For the cosmic X-ray and $\gamma$-ray diffuse background we assume the Gruber et al. \cite{gruber99} analytic form, derived from HEAO-1~A4 and COMPTEL measurements, and fitted in the energy range between 3 keV and 100 GeV. This spectrum is usually used as a standard reference \cite{frontera07,ajello08}, and is plotted in Figure~\ref{f:cxb}.

The first branch is valid below 60 keV:
\begin{equation}
F(E) = 7.877 \times \left( \frac{E}{1 \mathrm{\, keV}} \right)^{-1.29} e^{-E/41.13} \mbox{\, photons cm$^{-2}$ s$^{-1}$ sr$^{-1}$ keV$^{-1}$}
\end{equation}

while for energies above 60 keV:
\begin{eqnarray}
F(E) & = 0.0004317  \times \left( \frac{E}{60 \mathrm{\, keV}} \right)^{-6.5} \nonumber\\ & +  0.0084\times \left( \frac{E}{60 \mathrm{\, keV}} \right)^{-2.58}  \\ & +   0.00048  \times \left( \frac{E}{60 \mathrm{\, keV}} \right)^{-2.05}   \nonumber\\   & \mbox{\, photons cm$^{-2}$ s$^{-1}$ sr$^{-1}$ keV$^{-1}$}\nonumber
\end{eqnarray}

Other measurements of the CXB have been performed by, e.g., BeppoSAX \cite{frontera07}, INTEGRAL \cite{churazov07,turler10} and BAT \cite{ajello08}. These measurements are consistent with the HEAO-1 results, albeit some of these seems to indicate a slightly higher normalization ($\sim$8\%) for the $>$10 keV spectrum.

\subsubsection{Albedo $\gamma$-ray background}
\begin{figure}[htbp]
\centering
\includegraphics[width=.9\textwidth]{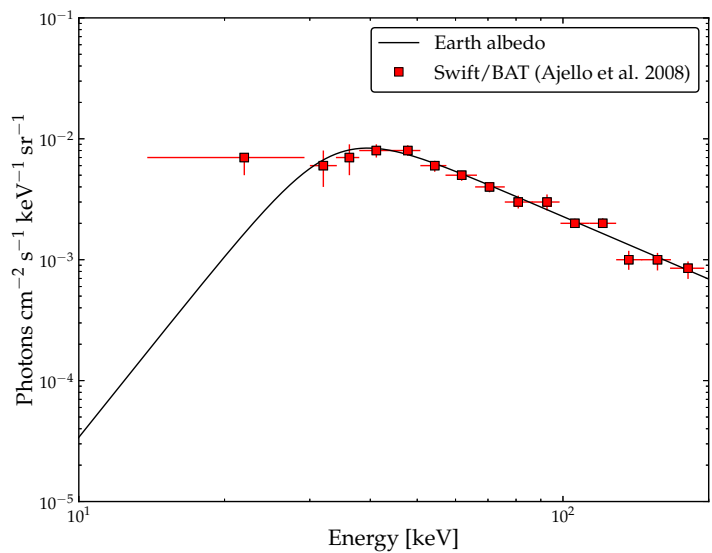}
\caption{Albedo $\gamma$-ray observations performed by BAT \cite{ajello08} and its parameterisation (Eq.~\ref{e:albedo}). }
\label{f:albedo}
\end{figure}

The secondary photon background is due to the interactions of cosmic rays (proton and leptonic components) with the atmosphere, and to the reflection of CXB emission. As such, it has a strong zenith dependence \cite{schoenfelder80}. 
This albedo component has a higher flux, for unit of solid angle, than the CXB for energies above $\sim$70 keV.
We assume the albedo spectrum as measured by BAT \cite{ajello08}, that agrees in the range above 50 keV, after some corrections, with previous measurements \cite{imhof76,dean89}. This spectrum can be parameterized by the function \cite{sazonov07,ajello08}:
\begin{equation}\label{e:albedo}
F(E) =   \frac{0.0148}{\left(\frac{E}{33.7 \mathrm{\, keV}}\right)^{-5} + \left(\frac{E}{33.7 \mathrm{\, keV}}\right)^{1.72}} \mbox{\, photons cm$^{-2}$ s$^{-1}$ sr$^{-1}$ keV$^{-1}$}
\end{equation}
and is plotted in Figure \ref{f:albedo}.

\subsection{Earth neutron albedo}

\begin{figure}[htbp]
\centering
\includegraphics[width=0.9\textwidth]{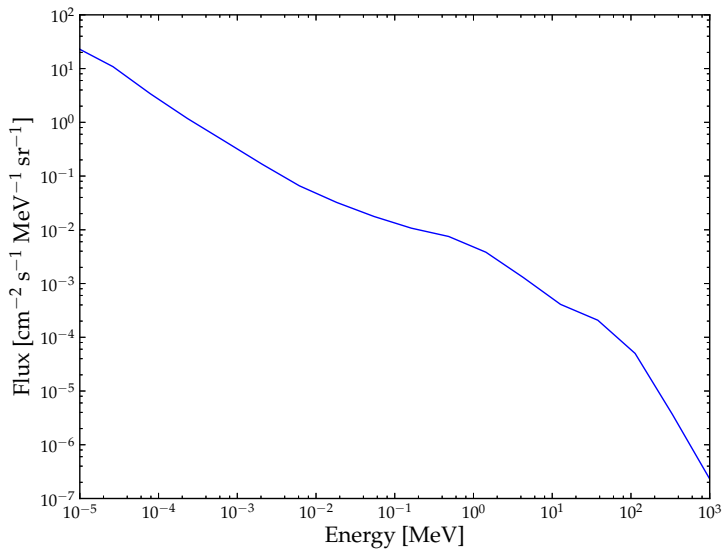}
\caption{Model spectrum of the Earth albedo neutrons for the LOFT orbit.}
\label{f:neutrons}
\end{figure}

As reported by the official ESA ECSS documents\footnote{\url{http://space-env.esa.int/index.php/ECSS-10-4.html}}, there is presently no model for atmospheric albedo neutron fluxes considered mature enough to be used as a standard.
To account for the flux of neutrons produced by cosmic-ray interactions in the Earth atmosphere, we however used the results of the QinetiQ Atmospheric Radiation Model (QARM, \cite{lei04,lei06}), that uses a response function approach based on Monte Carlo radiation transport codes to generate directional fluxes of atmospheric secondary radiation. 
The resulting spectrum, shown in Figure \ref{f:neutrons}, is also consistent with the Monte Carlo simulations of Armstrong \& Colborn \cite{armstrong92,fioretti12}.

\subsection{Natural radioactivity}\label{naturalradioactivity}
The lead-glass used in the microcapillary plate collimators contains potassium, in a fraction of approximately 7.2\% by weight. The activity due to the naturally occurring long-lived radioactive isotope $^{40}$K (with an abundance of 0.0117\% with respect to natural potassium) is to be taken into account. This isotope decays with the emission of $\beta$ rays having a continuum of energies up to 1.31 MeV ($\sim$89\% branching ratio) and of 1.46 MeV photons ($\sim$11\% b.r.) due to electronic capture.
Let $L$, $\rho$, $f_\mathrm{OAR}$ be the thickness, density and open fraction of the collimator. Therefore, the collimator activity for unit area is:
\begin{equation}
 A_\mathrm{coll} = A_K \cdot M_K
\end{equation}
where $A_K \sim 30$ Bq/g is the specific activity of natural potassium due to $^{40}$K, and $M_K$ is the total potassium mass per unit area in the collimator, i.e.
\begin{equation}
M_K = L(1-f_\mathrm{OAR})f_K
\end{equation}
where $f_K$ is the fraction by weight of potassium in the collimator glass. For the MCP lead glass, we have $f_K \sim 7.2$\%. 
Assuming for the MCP collimator $L = 6$ mm, $\rho = 3.3$ g/cm$^3$, $f_\mathrm{OAR} = 69.44$\% we obtain $M_K =  436$ g/m$^2$.
Therefore, we have $A_\mathrm{coll} =  13100$ Bq/m$^2$. For a total geometrical LAD collimator area of $\sim$15 m$^2$ we have thus $\sim$196500 decays/s.
Considering the branching ratio, we have therefore that every second in the collimator bulk are uniformly and isotropically generated $\sim$175000 electrons with a continuum of energies and $\sim$21000 monochromatic photons with a 1.46 MeV energy.

\section{The GEANT-4 LOFT/LAD simulator}\label{s:massmod}
\subsection{The mass model}
Simulations were performed using the Geant-4 Monte Carlo toolkit \cite{agostinelli03} (version 4.9.4). Geant-4 allows to describe the geometry and the materials of the instruments and of the satellite bus. Moreover, the code enables to follow the various physical interactions along the path of a primary event through the various components of the geometry, evaluating the secondary particles and energy deposits generated.

For the description of the electromagnetic interactions,  the Low Energy Livermore library\footnote{\url{https://twiki.cern.ch/twiki/bin/view/Geant4/LoweMigratedLivermore}} is used, that trace the interactions of electrons and photons with matter down to about 250 eV, using interpolated data tables. 
Physical processes like photoelectric effect, Compton and Rayleigh scattering, pair production, brems\-strahlung, multiple scattering and annihilation are simulated, optionally taking into account also the effect of photon polarization. Fluorescence X-rays and Auger electrons from the various chemical elements are included.
The hadronic physics, instead, is described using different models in different energy ranges (e.g. Bertini Cascade model, Quark-Gluon String model, etc.\footnote{See the Geant-4 Physics Reference Manual for more details.}). Both elastic and inelastic processes are treated, the latter describing e.g. nuclear capture, fragmentation, de-excitation and scattering of the relevant stable and long-lived nucleons and mesons.

\begin{figure}[htbp]
\centering
\includegraphics[width=0.6\textwidth]{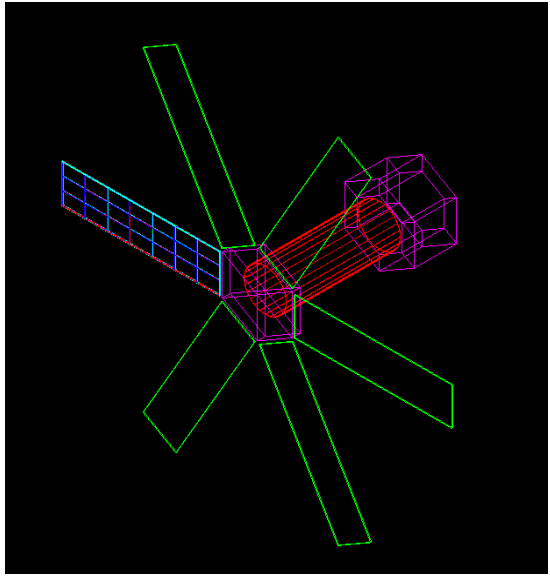}
\caption{The Geant-4 LOFT/LAD mass model.}
\label{f:massmodel}
\end{figure}

\begin{figure}[htbp]
\centering
\includegraphics[width=\textwidth]{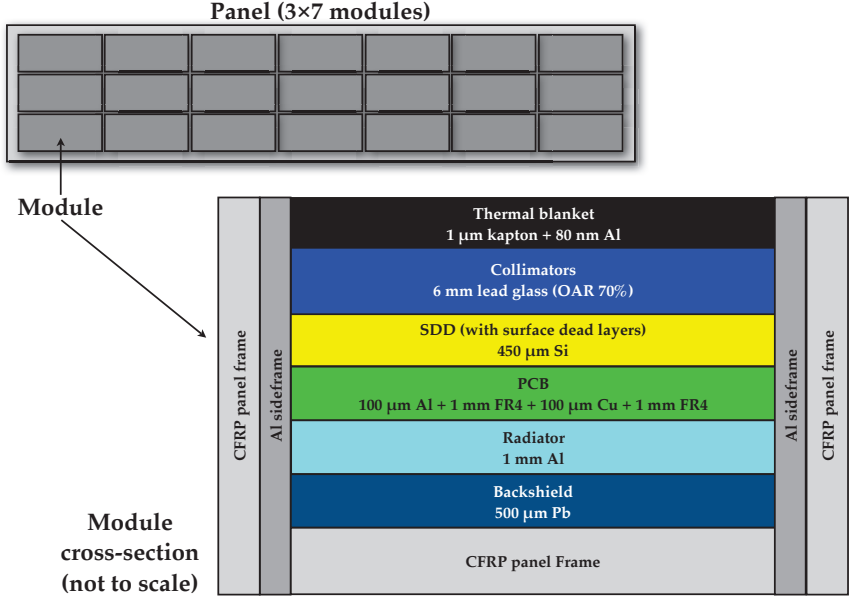}
\caption{Sketch of the simulated Geant-4 LOFT/LAD module geometry.}
\label{f:LAD_sketch}
\end{figure}

In Figure~\ref{f:massmodel} is shown the mass model geometry used in the simulations for the LAD instrument. 
A single LAD panel is simulated, with the actual dimensions ($\sim$350 cm  $\times$  90 cm) but with a simplified design that consists of a stacked-layers geometry (sketched in Figure~\ref{f:LAD_sketch}).
A simplified geometry for the satellite bus, the structural tower and the other five panels is assumed, using aluminium as the material with an ``effective'' density that takes into account the total mass and volume. 

The SDD array is represented as a 450 $\mu$m thick slab of silicon, subdivided in 0.97 mm  $\times$  35 mm pixels. On its surface are placed various passive layers (Al cathode implants, SiO$_2$ passivation layer, undepleted Silicon bulk), the MCP collimator and the optical/thermal filter (1 $\mu$m kapton coupled with a 80 nm Al layer). 
The MCP collimator has an ``effective'', reduced density, to take into account the holes without resorting to a more accurate but much more time-consuming geometrical description. 
Simulations on a smaller geometrical size have been performed to compare the ``effective'' collimator with a complete geometrical capillary plate description (see Sect. \ref{internal40Kactivity}), giving consistent results.
Underneath the detectors, the FEE board is simulated as a 2 mm thick FR4 slab with an interposed 100 $\mu$m Cu conductive layer and a 100 $\mu$m Al shielding layer, a 1 mm thick Al radiator and a 500 $\mu$m Pb backshield. 
An aluminium frame encloses the module sides, while the panel structure consists of a carbon-fiber reinforced plastic grid frame.

\subsection{Primary event generation}
For the generation of primary events, the General Particle Source\footnote{\url{http://reat.space.qinetiq.com/gps/}.} (GPS) approach is followed, that allows to produce arbitrary spectra in rather complex geometries. 
The initial events (particles or photons) are generated on the surface of a sphere of radius $R$, with a cosine-biased emission angle to ensure an isotropic flux. The emission angle is further restricted  between 0 (normal to the spherical surface) to $\theta_\mathrm{max}$: the emission cone then subtends a smaller sphere of radius $r$ that surrounds the experiment. We have:
\begin{equation}
\theta_\mathrm{max} = \arctan\left(\frac{r}{R}\right).
\end{equation}

Let $\Phi$ be the energy-integrated flux, between $E_\mathrm{min}$ and $E_\mathrm{max}$, expressed in particles cm$^{-2}$ s$^{-1}$ sr$^{-1}$.
The total rate is therefore:
\begin{equation}
N_{r} = \Phi 4 \pi^{2} R^{2} \sin^{2}\theta_\mathrm{max}
\end{equation}
This derives from the integration over the $2\pi$ emission angle for a point on the spherical surface, biased with the cosine law:
\begin{equation}
\int_{0}^{2\pi} d\phi \int_{0}^{\pi/2} d\theta \cos\theta\sin\theta = \pi
\end{equation}
and integrated over the source sphere surface $S = 4\pi R^{2}$. The restriction on the emission angle introduces the further factor $\sin^{2}\theta_\mathrm{max}$.

Therefore the simulation time that corresponds to $N$ generated primary events is:
\begin{equation}
\tau = \frac{N}{\Phi 4 \pi^{2} R^{2} \sin^{2}\theta_\mathrm{max}}
\end{equation}

If we detect $C_{i}$ counts in the energy bin $i$, the corresponding \emph{measured flux} (in counts cm$^{-2}$ s$^{-1}$), i.e. convolved with the detector response, would be:
\begin{equation}
F_{i} = \frac{C_{i}}{\tau A_\mathrm{det}}
\end{equation}
where $A_\mathrm{det}$ is the detector sensitive area.

\subsection{Anode multiplicity rejection filter}\label{amrf}
\begin{figure}[htbp]
\centering
\includegraphics[width=\textwidth]{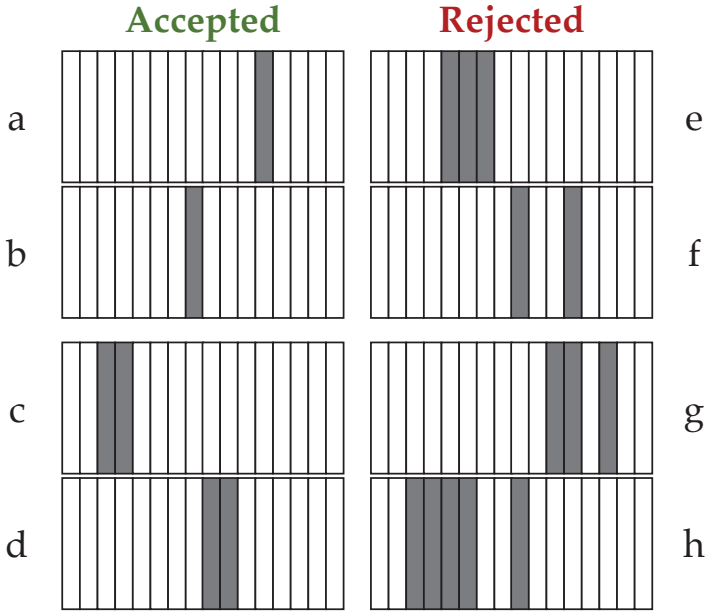}
\caption{The anode multiplicity rejection filter. Four SDDs are conceptually shown, each subdivided into two independent halves. Darker rectangles represent triggering anodes. Events $a$ and $b$ are single-anode energy deposits, while $c$ and $d$ are double-anode events: all of these are accepted by the filter. On the contrary, triple-anode events ($e$) or disconnected multiple-anode events ($f, g, h$) are rejected.}
\label{multiplicity_filter}
\end{figure}

For the LAD, an anode multiplicity rejection algorithm is implemented to filter out ionization streaks from charged particles, that leave an energy deposit on more than 2 adjacent anodes.
More in detail, for each of the two independent halves of a SDD tile, events are rejected if they trigger more than two adjacent anodes. Likewise, events are rejected if there is a group of non-adjacent triggering anodes within a ``distance" of 32 channels in the same half-SDD (Figure \ref{multiplicity_filter}). Energy deposits at a further distance are treated as independent.

Since the average energy loss of minimum ionizing particles (MIPs) in silicon is about 3.7~MeV/cm, their total energy deposit is usually above $\sim$150~keV. 
The anode multiplicity rejection filter, combined together with an upper threshold on the reconstructed event signal, is very effective in the suppression of the particle-induced background: a threshold of 80~keV allow to filter out 94\%--96\% of the particle background, depending on the incoming particle type and spectrum.
Of course, since this event elaboration is performed in the front-end electronics, dead-time calculations should be performed using the total incoming background rate.
The acceptance efficiency of this filter, defined as the fraction of photons from a ``true'', on-axis X-ray source that survive the multiplicity rejection, is very high. For a Crab-like spectrum, the efficiency is 99.98\%.

\section{Simulation results}\label{s:results}

The simulations have been performed by generating an isotropic flux of primary events (photons or particles, the latter evaluated at the Solar minimum, the worst case condition) on a sphere surrounding the experiment,  recording the energy deposits in the detector pixels and then applying the proper filter for the event multiplicity (Sect.~\ref{amrf}). The resulting counts have been then properly normalized, taking into account the energy spectra and solid angles of the simulated background contributions.

Before discussing the general results for the LAD background and a preliminary assessment of its stability, in the next two subsections we deal with some subleties that required a more detailed geometrical description and event processing. 

\subsection{Capillary reflections}
\begin{figure}[htbp]
\centering
\includegraphics[width=\textwidth]{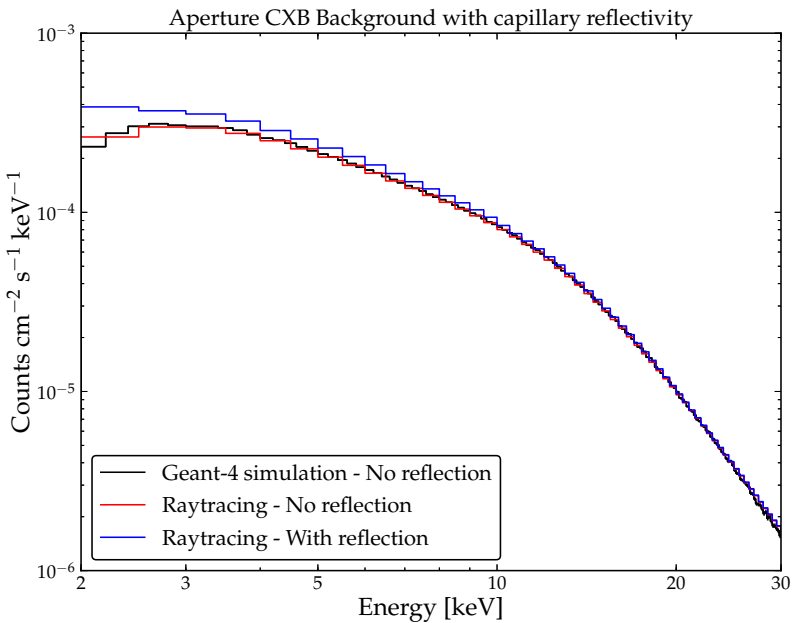}
\caption{Comparison of results between the Geant-4 code and a stand-alone raytracing simulator developed to evaluate the effects of the reflection of grazing-incident photons on the inner walls of the collimator.}
\label{reflection}
\end{figure}

A standalone, more detailed raytracing simulator has been built for the MCP to evaluate the effects of the grazing incidence angle reflectivity from the pore inner walls for the CXB photons collected in the $\sim 1^\circ \times 1^\circ$ field of view.
The code tracks photons from their incoming direction up to the detector plane modeling their interaction with the optical/thermal filter and the collimator structure. 
The simulated collimator has a full geometrical description, 6 mm thick filled with square pores 100 $\mu$m wide and 20 $\mu$m thick walls. 
The absorption and transmission probability of the optical/thermal filter elements are computed from tables derived from the database of the National Institute of Standards and Technology\footnote{\url{http://www.nist.gov}}; the lead glass attenuation coefficient is derived from tabulated values (G. Fraser, priv. comm.). Reflectivity from the pore walls, as derived from a set of laboratory measurements (G. Fraser, priv. comm.) is well modeled by data derived from CXRO database\footnote{\url{http://henke.lbl.gov/optical_constants}} corresponding to lead glass with a 11.8 nm surface micro-roughness value.
Results from this code, when reflection is not taken into account, are well in agreement with those obtained by the Geant-4 simulator (Figure \ref{reflection}).
The effect of the wall reflectivity produces an increase of the aperture CXB (Mineo et al., in preparation), going from 40\% at energies lower than 3 keV, down to $\sim$10\% between 5 and 10 keV and below 5\% above 10 keV.

\subsection{Internal $^{40}$K activity}\label{internal40Kactivity}
Not all the potassium decay products (see Sect. \ref{naturalradioactivity}) will leave a signal in the detector: the emission is isotropic, and the photon (or the electron) has to cross different thicknesses of lead glass, at various pitch angles, before reaching the SDD. Moreover, the efficiency of the silicon detector should be taken into account; eventually, only a fraction of the detected events will have a pulse height amplitude corresponding to the standard LAD energy band (2--30 keV).

Primary particles (1.46 MeV monochromatic photons and electrons having energies between 1 keV and 1311 keV) have been generated randomly in the lead-glass walls of a $\sim1.2\times1.2$ cm complete model of a MCP geometry and mass (6 mm thickness, 100 $\mu$m square holes, 20 $\mu$m wall thickness, $\sim$70\% OAR) coupled to a SDD detector.
It is found that the decay of one $^{40}$K atom generating a 1.46 MeV photon has a very small probability to generate a signal with equivalent amplitude between 2 and 30 keV, obtaining we obtain a background rate of only $\sim 5\times10^{-5}$ counts cm$^{-2}$ s$^{-1}$. Therefore, for all practical purposes, we can discard the effect of the $^{40}$K electronic capture decay channel.
Electrons produced by the $\beta$ decay, on the contrary, have a higher chance to cause a signal, 
producing a background rate of $\sim 2\times10^{-3}$ counts cm$^{-2}$ s$^{-1}$.

\subsection{The LAD background}
\begin{figure}[htbp]
\centering
\includegraphics[width=\textwidth]{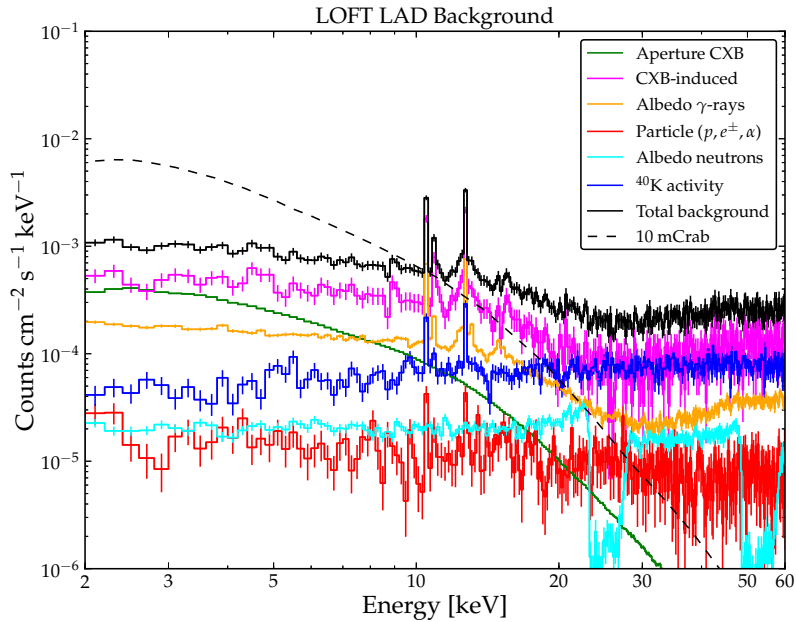}
\caption{The LAD total background and its various components discussed in the text. The spectrum corresponding to a 10 mCrab source is shown as a dashed line. Note that the latter spectrum and the aperture CXB one have been normalized relatively to the other curves according to the reduced effective area that includes the collimator OAR (i.e. $\sim$10 m$^2$ at 8~keV).}
\label{LADbkg}
\end{figure}

The total resulting LAD background is shown in Figure \ref{LADbkg}, while the breakdown of the count rate in its various components is shown in Table~\ref{LADbkgtable}.
No detector energy resolution smoothing of the spectra was included here.

The main background contribution is due to the high-energy photons from the diffuse cosmic X-ray background and the Earth albedo that leaks from and are scattered in the collimators. These two components alone accounts for about 70\% of the background count rate in the 2--30 keV band.

The diffuse emission collected in the field of view (even including the effect of capillary reflectivity), the particle-induced and internal backgrounds are a smaller contribution to the total count rate.
While the CXB emission collected in the aperture field of view has a significant contribution only below $\sim$10 keV, cosmic-ray particles and neutrons produce almost flat spectra, similarly to the internal activity background, and these components become dominant only above $\sim$20~keV.
The dips in the neutron-induced spectrum are due to inelastic scattering resonances.

Fluorescence emission from the Pb contained in the collimator glass ($L$-shell lines at 10.55 and 12.61 keV) is well apparent, artificially emphasized by the non-inclusion of the energy resolution in the plot.
Analysis are ongoing to evaluate whether these lines are to be shielded or used as in-flight calibration lines.
In the plot is also shown, as a reference to the LAD science requirement, the spectrum of a 10 mCrab point-like source (dashed line).
The total background count rate in the 2--30 keV band corresponds to a flux of $\sim$9 mCrab, thus ensuring the fulfillment of the scientific requirements, and it is below 5 mCrab in the most important 2--10~keV band.

We can estimate a margin of error on the background rate, taking into account that the CXB emission and the Earth albedo fluxes are affected by a maximum error of $\sim$10\%--20\% on their normalization \cite{ajello08,turler10}.
The particle components are more uncertain \cite{mizuno04,alcaraz00a} (see also Sect.~\ref{s:primarycr}). However, these have been conservatively assumed at the Solar minimum, while LOFT, if selected, will fly around the next Solar maximum ($\sim$2024), when the particle flux is expected to be approximately 20\% less intense. We  consider a $\sim$50\% error on these fluxes. 
Weighting the above uncertainties on the LAD background components, we thus estimate an overall conservative maximum margin of error on the total background rate of $\sim$20\%, given the present geometrical model. Future developments in the mission design will allow for a more accurate mass model and consequently to refine these results.
Therefore, we can conclude that the LOFT scientific requirements are expected to be met.

\begin{table}[htdp]
\caption{The LAD background contributions. The last line shows the LAD requirement for the total background level in the 2--30 keV band.}
\begin{center}
\begin{tabular}{c|c|c|c|c}
Contribution & counts cm$^{-2}$ s$^{-1}$ & Percentage & mCrab & mCrab\\
& 2--30 keV & 2--30 keV & 2--30 keV & 2--10 keV\\
\hline \hline
Aperture CXB   & 1.9 $\times$ 10$^{-3}$ & 13\%& 0.8 & 0.9\\
CXB-induced  & 7.4 $\times$ 10$^{-3}$ & 51\%& 4.6 & 2.4\\
Earth albedo $\gamma$-rays   & 2.8 $\times$ 10$^{-3}$ & 19\% & 1.7 & 0.8\\
Earth albedo neutrons   & 5.0 $\times$ 10$^{-4}$ & 2.5\% & 0.2 & 0.1\\
Cosmic-ray particles   & 3.6 $\times$ 10$^{-4}$ & 3.5\% & 0.3 & 0.08\\ 
$^{40}$K activity   & 2.0 $\times$ 10$^{-3}$ & 14\% & 1.2 & 0.3\\ \hline
Total background   & 1.4 $\times$ 10$^{-2}$ & 100\% & 8.8 & 4.6\\ \hline
Requirement   & 1.6 $\times$ 10$^{-2}$ &--- & 10 & --- \\
\end{tabular}
\end{center}
\label{LADbkgtable}
\end{table}
 
 \subsection{Background variability}\label{bkgvariation}
 
The main components of the LAD background are shown in Figure \ref{LADbkg}.
The high energy photons from CXB emission and the Earth $\gamma$-ray albedo leaking through the collimator 
represent the dominant background component. Although they are intrinsically steady and predictable, due to their different intensity and 
spectra a varying relative orientation of the LAD in their radiation environment will cause
a small and smooth modulation of the detected background, on the orbital
timescales ($\sim$90 minutes). This is due to the varying viewing geometry along the orbit and for
different attitudes. This effect has been studied through simulations, finding that the maximum expected
modulation of the background is less than 10\%. 
This value has to be compared to a factor
of a few for instruments dominated by particle-induced background. For example,
RXTE/PCA had up to a $\sim$250\% variation on orbital timescales \cite{jahoda06}. 
In the LOFT case, the effect of the other potentially varying sources, i.e. particle induced background,
is greatly reduced by the very stable environment offered by the low Earth equatorial orbit and by the fact 
that this component accounts for less than 6\% of the overall background.

\begin{figure}[htbp]
\centering
\includegraphics[width=0.9\textwidth]{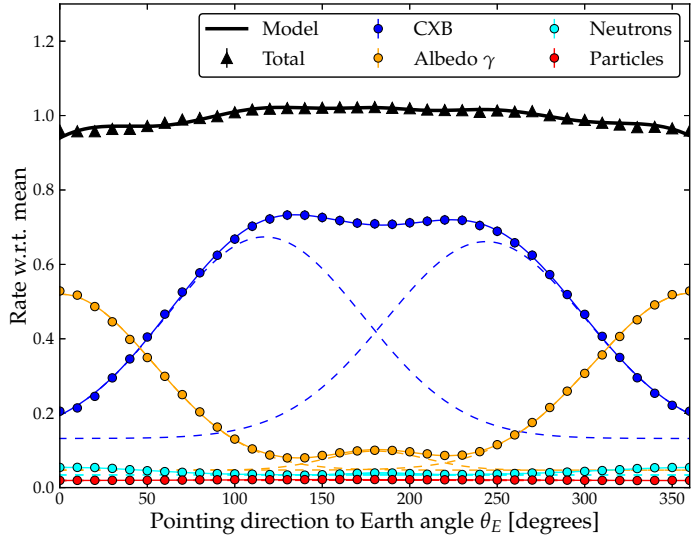}
\caption{Non-aperture background modulation due to the varying position of the Earth with respect to the pointing direction. $\theta_E$ is the polar angle between the LAD pointing axis and the center of the Earth. Blue points are the CXB-induced contribution, orange points the Earth $\gamma$-ray albedo, cyan points the albedo neutrons and red points the particles. The black triangles represent the total background, while the black continuous curve is the background model. All quantities have been scaled to the average isotropic background count rate in the 2--30~keV band.}
\label{bkg_rot_rate_avg}
\end{figure}

In Figure \ref{bkg_rot_rate_avg} the background rate is shown as a function of the angle between the 
LAD pointing direction and the center of the Earth. 
$\theta_E = 0^\circ$ represents the Earth center aligned with the field of view, 
while $\theta_E = 180^\circ$ stands for the Earth at the instrument nadir. 
In practice, this corresponds to the orbital modulation for a low-declination source. 
The curve representing the total background (black symbols) shows a maximum modulation of $\sim$8\%.
The modulation of the background rate is entirely due to geometrical effect and it can be predicted and modeled, thanks to the intrinsic steadiness of the relevant sources. Each background component is well modeled as a function of the Earth location with respect to the pointing direction by a sum of two Gaussian distributions,
centered at about $\theta_E = 0^\circ$ and 180$^\circ$ for the Earth-originated components and at about $\theta_E = 120^\circ$ and 240$^\circ$ for the diffuse sky components. These distributions, also reported in Figure \ref{bkg_rot_rate_avg}, arise from the convolution of the directional ``transparency'' of the LAD instrument with the Earth-occulted field of view. Of course, when the detector points towards the zenith ($\theta_E = 180^\circ$) the contribution from the leaking CXB emission is maximum. The overall convolution of these out-of-phase components is to give a very small fluctuation of the total background. Moreover, for an actual pointing towards an astrophysical source, the range of possible Earth angles $\theta_E$ is restricted, from a $\theta_E = 90^\circ$ for a source at the orbital pole (that is nearly coincident with the Celestial pole) to the full 0$^\circ$--180$^\circ$ range for equatorial sources. 
This lowers the background modulation further below the maximum during the observation of a realistic scientific target.
The geometrical model, properly calibrated using in-orbit flat fields, is anticipated to allow for a background prediction at a level significantly better than 1\% in the 2--10 keV band, which is the LAD science requirement. 
Such a level of systematic uncertainty was indeed already reached by the past experiment RXTE/PCA, which, in the presence of a much more variable background level (250\% vs 8\%) and less predictable background sources, with an appropriate modeling reached the level of $\sim$1\% \cite{jahoda06,shaposhnikov12}. 

However, as some of the LOFT science cases (in particular the extragalactic science) 
will benefit from reaching a background knowledge significantly better than the requirement, 
an active background monitoring was designed for the LAD, to further improve the modeling. 
Due to the slow and smooth background variation, there is no need for a high-statistics, instantaneous monitoring of the rate. 
Rather, a continuous benchmark of the slow modulation will allow for a real-time verification of the background model. 
This active background monitoring is achieved by the introduction of a ``blocked collimator'' (a collimator with the same stopping power but no holes) for an area corresponding to one Module of the LAD. This will enable the continuous monitor of all components of the LAD background, with the exception of the aperture background, accounting for $\sim$90\% of the total background, and also evaluating the long-term variations (e.g. linked to the Solar cycle). 
Preliminary simulations for different targets (i.e., attitude configurations) show that the accuracy in the background modeling during a typical observation can be improved down to $\sim$0.1\%--0.3\% by using these additional data.

The subject of time-dependent modeling of the background is being further studied in depth to definitely assess the ultimate instrument capability for weak sources. 
Here we just reported a few preliminary results allowing to identify the range of interest. An exhaustive report and discussion is deferred to a dedicated paper, currently in preparation.

\subsection{Absolute background level: cosmic variance and source contamination}\label{s:offaxissources}
In the previous section we preliminarily addressed the issue of background stability and modeling during a single observation. 
This is the key parameter for most of the science driver of observations of weak sources, 
where relative variations of source features will be studied (e.g., the Fe line in AGNs). 
Anyway, a few science cases may require an estimate of the absolute value of the background. 
This faces an intrinsic physical property of the CXB, known as \emph{cosmic variance} \cite{revnivtsev03}: 
the flux of the CXB is not perfectly isotropic but varies on different angular scales.
The variation was measured \cite{revnivtsev03,revnivtsev08,moretti09} as $\sim$7\% (rms) on the angular scale of 1$^\circ$ 
and $\sim$2\% on 20$^\circ$--40$^\circ$. 
In the case of the LAD, this affects only the aperture background, as the CXB/albedo contamination is seen from all directions. 
The expected variation in the LAD background due to the cosmic variance is given by the fraction of the total rate due to the aperture CXB (13\%) weighted by the variance over the field of view ($\sim$7\%), which is about 0.9\%. 
Given its intrinsic astrophysical nature, the only way of measuring this component is through a local blank field, just as for any other imaging or non-imaging experiment. This is indeed the plan for the few science cases requiring the absolute knowledge of the background rate rather than its stability. 
It is worth stressing that the cosmic variance does not affect at all the background stability in time for a given target observation.

A further small contribution to the ``local background'' (i.e., for a given target/attitude) is given by the possible contamination due to bright and hard sources outside the field of view. Similar to the leaking of diffuse CXB/albedo photons through the collimator, the same effect can occur to photons from point-like sources. 
We investigated this effect by simulations, and the maximum contribution to the overall background is of the order of 1--3\% (Crab-like sources, in 2--30 keV) down to 0.1\% (softer sources, e.g. Sco X-1). 
This additional minor contribution will be monitored by the active background monitoring system, as well as by the Wide Field Monitor. The quantitative assessment of the small effect on the background stability when the contaminating sources are variable on the time scale of interest will be extensively addressed in the paper in preparation mentioned earlier.

\section{Conclusions}\label{s:conclusions}

LOFT will be an innovative mission that will observe compact Galactic and extragalactic objects in both the spectral and the temporal domains.
The unprecedented sensitive area of the LAD instrument will open new windows in the study of the fundamental physics allowed by these natural laboratories.
The scientific objectives of the mission require an accurate knowledge and minimisation of the detector background.
To this end, an extensive mass-model for the LOFT/LAD experiment has been developed, using the standard Geant-4 toolkit (Sect.~\ref{s:massmod}), 
and all the main components (photonic, leptonic, hadronic and internal, Sect.~\ref{s:bkgsources}) of the orbital background environment have been simulated.

The main contribution to the overall background is found to be due
to the diffuse X-rays that ``leaks" from the lead-glass collimators of the instrument. 
This emission originates from the diffuse cosmic X-ray background and from the Earth albedo.
The particle-induced background is minimised mainly thanks to the LOFT low-inclination, low-altitude orbit and small mass for unit area of the LAD experiment,
becoming dominant only at high energies, above 30~keV.
A further suppression of the particle background is enabled by the particular signature that these events leave on the Silicon Drift Detectors (Sect.~\ref{amrf}).
The simulations show the feasibility for the current LAD instrument design to fulfill the required background level of 10 mCrab in the 2--30~keV band.

Background variations on orbital timescales are mostly induced by the varying geometry between the position of the Earth and the pointing direction,
As such, they can be modelled and accounted for.
The use of special detector units (``blocked'' module) 
and a carefully planned blank-field pointing strategy, 
are foreseen to monitor these background variations 
enabling to reach a residual systematics level better than 0.3\%.
The residual contribution of strong (and variable) off-axis sources, above 30--50 keV, if needed,
can be further modelled using observations performed with the other instrument onboard LOFT, the WFM.

An increase in the accuracy of the background determination for the LAD will be provided by further improvements in the geometrical description of the instrument itself and of the spacecraft bus structures.
Furthermore, results will be refined by taking into account a complete determination of the residual activation of the detector materials following the grazing passages in the high-background environment of the South Atlantic Anomaly, where however no scientific observations are conducted.

\begin{acknowledgements}
We are grateful to the anonymous referee whose helpful comments greatly improved the text, and to Alessandra De Rosa for useful discussions.
We acknowledge financial support from ASI/INAF contract I/021/12/0. RC furthermore acknowledge support from INAF Tecno-PRIN 2009 grant.
\end{acknowledgements}

\bibliographystyle{spphys}      
\bibliography{bibliography}   

\end{document}